\documentstyle[12pt]{article}
\def\be{\begin{equation}}
\def\ee{\end{equation}}

\begin{document}

\title{Interaction of 3--level atom with radiation}

\author{L.Accardi, K.Imafuku, S.V.Kozyrev}
\maketitle

\begin{abstract}
The interaction of 3--level system with a quantum field
in a non--equilibrium state is considered.
We describe a class of states of the quantum field for wich
a stationary state drives the system to inverse populated state.
We find that the quotient of the population of the energy levels
in the simplest case is described by the double Einstein formula
which involves products of two Einstein emission/absorption relations.
Emission and absorption of radiation by 3--level atom
in non--equlibrium stationary state is described.
\end{abstract}

\section{Introduction}

In the present paper we consider a 3--level system
(for example an atom) interacting with radiation.
In the work \cite{Kentaro} we apply the technique developed here
to describe stimulated emission for 3--level atom
interacting with radiation in a non--equilibrium state.
In the work \cite{spinboson} a 2--level quantum system interacting
with quantum field
was considered.
In  \cite{notes} a stationary nonequilibrium state for an
$n$--level system interacting with a nonequilibrium quantum field
was obtained.

We show that for such a system
application of the stochastic limit allows to obtain an interesting
effect of inversion of population: for a special choice of the state
of reservoir the system relaxes to stationary state where
the population of the level of the system with higher energy will be larger
than the population of the level with lower energy.

We obtain an equation for the number of photons
emitted and absorbed by the system.

We investigate the examples of 2--level and 3--level systems.
We show that for a 2--level atom the emission in the stationary
regime equals to absorption. For a 3--level system
emission and absorption of radiation are controlled by the state
of the field. We find that for a 3--level
system in a stationary nonequilibrium state two regimes are possible:
the emission and the absorption regime. In the emission regime
the total number of quanta in the system increases, and
in the absorption regime it decreases. These regimes are controlled
by the function $\beta(\omega)$ in (\ref{e_state}).
For example a 3--level system with energy levels
$\varepsilon_1<\varepsilon_2<\varepsilon_3$
is in emission regime when
$$
\beta(\varepsilon_2-\varepsilon_1)+
\beta(\varepsilon_3-\varepsilon_2)>\beta(\varepsilon_3-\varepsilon_1)
$$
and is in absorption regime when the opposite inequality holds.

In the emission regime the 3--level system converts
radiation with the frequency $\omega_2$ into radiation with
the frequencies $\omega_1$ and $\omega_3$ and vice versa in the
absorption regime.

This means that this stationary state of the 3--level system
gives an example of dissipative structure in
the Prigogine sense \cite{Prigogine}.

For an equilibrium state of the field, i.e. when the function $\beta(\omega)$
is linear, the system is in equilibrium with radiation.

The interaction of a quantum system with a quantum field is described by
an Hamiltonian
\begin{equation}\label{H}
H=H_S+H_R+\lambda H_I
\end{equation}
The system degrees of freedom are described by the system
Hamiltonian $H_S$.

The radiation degrees of freedom are described by the
Hamiltonian
\begin{equation}\label{e_reser}
H_R=\int \omega(k)a^*(k)a(k) dk
\end{equation}
where $a(k)$ is a bosonic field with a Gaussian state of the
following form
\begin{equation}\label{e_state}
\langle a^*(k)a(k')\rangle=N(k)\delta(k-k')= {1\over
e^{\beta(\omega(k))}-1} \delta(k-k')
\end{equation}
and $\beta(\omega(k))$ is a (non necessarily linear) function.

The interaction Hamiltonian $H_I$ is defined as follows
\begin{equation}\label{e_inter}
H_I=\int \overline{g(k)} a(k)D^* dk +\hbox{  h.c. }
\end{equation}

We investigate the dynamics of this system in the
stochastic limit, cf. \cite{book}, in the regime of
weak coupling ($\lambda\to 0$) and large times.
This regime is given by time rescaling to $t\mapsto {t\over \lambda^2}$.
This rescaling and the interaction (\ref{e_inter}) lead naturally
to introduce the rescaled quantum fields
\begin{equation}\label{5a}
{1\over \lambda} e^{-{it\over\lambda^2}(\omega(k)-\omega)}a(k)
\end{equation}
where $\omega$ are the Bohr frequencies (differencies of eigenvalues of
the system Hamiltonian $H_S$).

By the stochastic golden rule \cite{book}
the rescaled field (\ref{5a}) in the stochastic limit
becomes a quantum white noise $b_{\omega}(t,k)$, or master field
satisfying the commutation relations
\begin{equation}\label{2_5}
[b_{\omega}(t,k),b^*_{\omega'}(t',k')]=\delta_{\omega ,\omega '}
2\pi\delta(t-t')\delta(\omega(k)-\omega)\delta(k-k')
\end{equation}
and with the mean zero gauge invariant Gaussian state with correlations:
\begin{equation}\label{cormafi1}
\langle b^*_{\omega}(t,k)b_{\omega'}(t',k')\rangle=\delta_{\omega ,\omega'}
2\pi\delta(t-t')\delta(\omega(k)-\omega)\delta(k-k')N(k)
\end{equation}
\begin{equation}\label{cormafi2}
\langle b_{\omega}(t,k)b^*_{\omega'}(t',k')\rangle=\delta_{\omega ,\omega'}
2\pi\delta(t-t')\delta(\omega(k)-\omega)\delta(k-k')(N(k)+1)
\end{equation}
White noises, corresponding to different frequencies $\omega$,
are independent.

The Schr\"odinger equation becomes a white noise Hamiltonian equation,
cf. \cite{book}, \cite{notes} which when put in normal order is
equivalent to the quantum stochastic differential equation (QSDE)
\begin{equation}\label{1.21}
dU_t=(-idH(t)-Gdt)U_t \qquad ;\quad t>0
\end{equation}
with initial condition $U_0=1$ and where

(i) $h(t)$ is the white noise Hamiltonian and $dH(t)$,
called {\it the martingale term\/}, is the stochastic differential:
\begin{equation}\label{2_9}
dH(t)=\int^{t+dt}_th(s)ds=
\sum_{\omega}\left(
E_{\omega}^*\left(D\right)dB_{\omega}(t)+
E_{\omega}\left(D\right)dB^*_{\omega}(t)
\right)
\end{equation}
driven by the quantum Brownian motions
\begin{equation}\label{2_10}
dB_{\omega}(t):=\int^{t+dt}_t\int dk\overline g(k)b_{\omega}(\tau,k)d\tau
=:\int^{t+dt}_t b_{\omega}(\tau,g)d\tau
\end{equation}

(ii) The operator $G$, called the {\it drift\/}, is given by
\begin{equation}\label{drift1}
G=\sum_{\omega}
\left(
(g|g)^-_{\omega}
E_{\omega}^*\left(D\right)E_{\omega}\left(D\right)
+\overline{(g|g)}^+_{\omega}
E_{\omega}\left(D\right)E_{\omega}^*\left(D\right)
\right)
\end{equation}
where the explicit form of the constants $(g|g)^{\pm}_{\omega}$,
called the generalized susceptivities, is:
\begin{equation}\label{1.27a}
(g|g)^{-}_{\omega}=
\int dk\, |g(k)|^2
{-i(N(k)+1)\over\omega(k)-\omega-i0}=
\end{equation}
$$
=\pi\int dk\, |g(k)|^2 (N(k)+1) \delta(\omega(k)-\omega)
-i\,\hbox{P.P.}\,\int dk\, |g(k)|^2
{(N(k)+1)\over\omega(k)-\omega}$$
\begin{equation}\label{1.27b}
(g|g)^{+}_{\omega}=\int dk\, |g(k)|^2
{-iN(k)\over\omega(k)-\omega-i0}=
\end{equation}
$$
=\pi\int dk\, |g(k)|^2 N(k) \delta(\omega(k)-\omega)
-i\,\hbox{P.P.}\,\int dk\, |g(k)|^2
{N(k)\over\omega(k)-\omega}
$$

In the present paper we consider a generic quantum system,
for which for each Bohr frequency $\omega$
there exist a unique pair of eigenstates
$|1_{\omega}\rangle$ and $|2_{\omega}\rangle$
corresponding to the two energy levels, $\varepsilon_{1_\omega},
\varepsilon_{2_\omega}$, so that
$$
\omega=  \varepsilon_{2_\omega}-\varepsilon_{1_\omega}
$$
In this case
\begin{equation}\label{1.29c}
E_{\omega}(D)=
\langle 1_{\omega}| D |2_{\omega}\rangle |1_{\omega}\rangle\langle 2_{\omega}|
\end{equation}
$$
E_{\omega}(D)E^*_{\omega}(D)=
|\langle 1_{\omega}| D |2_{\omega}\rangle|^2
|1_{\omega}\rangle\langle 1_{\omega}|
$$
$$
E^*_{\omega}(D)E_{\omega}(D)=
|\langle 1_{\omega}| D |2_{\omega}\rangle|^2
|2_{\omega}\rangle\langle 2_{\omega}|
$$

We consider a dispersion $\omega(k)$ which is $\ge 0$ and,
moreover, we suppose that the Lebesgue measure of the set
$\{k:\omega(k)=0\}$ equals to zero.
This implies that the real part of generalized
susceptivities $\hbox{ Re }(g|g)^\pm_{\omega}$ is non--negative
and can be non--zero only for $\omega>0$.

We will also use the notation $(g|g)^\pm_{ij}$ for
$(g|g)^\pm_{\omega}$ if $\omega=\varepsilon_i-\varepsilon_j$.

In the present paper we investigate the non--eqilibrium stationary states
for the master equation satisfied by the diagonal part of
the density matrix of a generic quantum system.
This equation was obtained in \cite{notes} and is
\begin{equation}\label{diagevol}
{d\over dt}\,\rho(\sigma,t)=\sum_{\sigma':\varepsilon_{\sigma'}>
\varepsilon_\sigma}(\rho(\sigma',t)2\hbox{Re
}(g|g)^-_{\sigma'\sigma}- \rho(\sigma,t)2\hbox{ Re
}(g|g)^+_{\sigma'\sigma})|\langle\sigma,D \sigma'\rangle|^2+
$$
$$
+\sum_{\sigma':\varepsilon_\sigma>\varepsilon_{\sigma'}}(\rho(\sigma',t)
2\hbox{Re }(g|g)^+_{\sigma\sigma'}-\rho(\sigma,t)2\hbox{Re
}(g|g)^-_{\sigma\sigma'})|\langle\sigma',D\sigma\rangle|^2
\end{equation}
where $\rho(\sigma,t)=\rho(\sigma,\sigma,t)$ and
$|\sigma\rangle$ are eigenvectors of the system Hamiltonian $H_S$.

If the system has a finite
number of energy levels, there exists a stationary state for the
evolution driven by the above master equation and, if the state of the
reservoir is non--equilibrium, then the stationary state does not
satisfy the detailed balance condition for the master equation.

The diagonal and the off--diagonal terms of the
reduced density matrix evolve separately.
The off--diagonal part of the density matrix evolves independently
and vanishes exponentially, cf. \cite{notes}. This corresponds to
the collapse of a quantum state to a classical mixed state,
described by the diagonal part of the density matrix. The diagonal
part $\rho(\sigma,\sigma,t)$ may be considered as a classical
distribution function and equation (\ref{diagevol}) may be
considered as a kinetic equation it.

The structure of the present paper is as follows.

In section 2 we describe the stationary state for a 3--level atom
interacting with radiation found in \cite{notes}.

In section 3 we investigate the properties of this stationary
and find that the quotient of the populations
of energy levels in the considered nonequilibrium stationary state
does not satisfy the Einstein emission/absorption relation.
We find that in particular case the quotient of populations
will satisfy a new relation that we call the Double Einstein relation.

In section 4 we use the form of this state to describe
the inversion of population in our 3--level system.

In section 5 we derive a master equation for the density of photons.

In section 6 we use this equation to investigate emission and absorption
of radiation by the system in the non--equilibrium stationary state,
obtained in \cite{notes}.

\section{Stationary state for 3--level system}

Consider a 3--level system with energy levels $|1\rangle$,
$|2\rangle$, $|3\rangle$ with energies
$\varepsilon_1<\varepsilon_2<\varepsilon_3$
and Bohr frequencies
$$
\omega_1=\varepsilon_2-\varepsilon_1,\quad
\omega_2=\varepsilon_3-\varepsilon_1,\quad
\omega_3=\varepsilon_3-\varepsilon_2
$$
In the work \cite{notes} we found that the dynamics
generated by the master equation (equation for reduced
density matrix of the system) describes relaxation of the system
to a stationary state.
In this state the off--diagonal (in the basis of egenvectors
of the system Hamiltonian $H_S$) elements of the reduced density matrix
are equal to zero (this describes the effect of decoherence of a system
coupled to reservoir) and the diagonal elements of the reduced density
matrix for 3--level system take the form:
\begin{equation}\label{x}
\rho_1=|\langle 1D2\rangle|^2|\langle 1D3\rangle|^2{I(\omega_1)\over
1-e^{-\beta(\omega_1)}} {I(\omega_2)\over
1-e^{-\beta(\omega_2)}}+|\langle 1D2\rangle|^2|\langle
2D3\rangle|^2{I(\omega_1)\over
1-e^{-\beta(\omega_1)}}{I(\omega_3)\over 1-e^{-\beta(\omega_3)}}+
$$
$$
+|\langle 1D3\rangle|^2|\langle 2D3\rangle|^2{I(\omega_2)\over
1-e^{-\beta(\omega_2)}}{I(\omega_3)\over e^{\beta(\omega_3)}-1}
\end{equation}
\begin{equation}\label{y}
\rho_2= |\langle 1D2\rangle|^2|\langle 1D3\rangle|^2{I(\omega_1)\over
e^{\beta(\omega_1)}-1} {I(\omega_2)\over
1-e^{-\beta(\omega_2)}}+|\langle 1D2\rangle|^2|\langle
2D3\rangle|^2{I(\omega_1)\over
e^{\beta(\omega_1)}-1}{I(\omega_3)\over 1-e^{-\beta(\omega_3)}}+
$$
$$+
|\langle 1D3\rangle|^2|\langle 2D3\rangle|^2{I(\omega_2)\over
e^{\beta(\omega_2)}-1}{I(\omega_3)\over 1-e^{-\beta(\omega_3)}}
\end{equation}
\begin{equation}\label{z}
\rho_3= |\langle 1D2\rangle|^2|\langle 1D3\rangle|^2{I(\omega_1)\over
1-e^{-\beta(\omega_1)}}{I(\omega_2)\over
e^{\beta(\omega_2)}-1}+|\langle 1D2\rangle|^2|\langle
2D3\rangle|^2{I(\omega_1)\over
e^{\beta(\omega_1)}-1}{I(\omega_3)\over e^{\beta(\omega_3)}-1}+
$$
$$
+ |\langle 1D3\rangle|^2|\langle 2D3\rangle|^2{I(\omega_2)\over
e^{\beta(\omega_2)}-1} {I(\omega_3)\over e^{\beta(\omega_3)}-1}
\end{equation}
where $\rho_i=\langle i|\rho|i\rangle$ and
$$
I(\omega)=\int |g(k)|^2 \delta(\omega(k)-\omega)dk
$$

\section{The Double Einstein formula}

Consider the quotient
\begin{equation}\label{emabs}
{\hbox{Re }(g|g)^{-}_{\omega}\over \hbox{Re }(g|g)^{+}_{\omega}}=
{N(\omega)+1\over N(\omega)}
\end{equation}

Recalling that $N(\omega)$ is the density of field quanta
(photon, phonons,\dots) at frequency $\omega$, and comparing formula
(\ref{emabs}) with the well known formula of radiation theory
(the Einstein formula)
\begin{equation}\label{heitl18}
{W_{\hbox{emission}}\over W_{\hbox{absorption}}}=
{\overline{n}_\omega+1\over \overline{n}_\omega}
\end{equation}
giving the quotient of the probability of emission and absorption
of a light quantum by an atom (cf. \cite{Heitler}, Chap. V, paragraph 17,
formula (18)), we gain some physical intuition of the meaning
of the generalized susceptivities. In fact the quotient (\ref{heitl18})
{\it \dots is just that which is necessary to preserve the correct
thermal equilibrium of the radiation with the gas \dots}
(\cite{Heitler}, p.180).

In the stochastic limit approach
this statement can be proved using the master equation (\ref{diagevol})
for the diagonal part of density matrix.
One can prove that, if the state of reservoir is equilibrium, then
the dynamics generated by the master equation describes the relaxation
to equilibrium state of the system satisfying the detailed balance condition
for (\ref{diagevol}) i.e.:
\begin{equation}\label{rhooverrho}
{\rho_{\sigma'}\over \rho_{\sigma}}=
{\hbox{Re }(g|g)^{-}_{\omega}\over \hbox{Re }(g|g)^{+}_{\omega}}=
{N(\omega)+1\over N(\omega)},\qquad
\omega=\varepsilon_{\sigma} - \varepsilon_{\sigma'}>0
\end{equation}
For equilibrium state the quotient of populations
of the two levels with energy difference $\omega$
is equal to the Einstein emission--absorption quotient for quanta
with energy $\omega$.
This suggests that the quotients (\ref{emabs})
may play a similar role for some stationary non equilibrium states.

Let us give an example of such a state for which we will get
a generalization of condition (\ref{rhooverrho}).
Consider the state (\ref{x}), (\ref{y}), (\ref{z}).
For simplicity we consider the case when the matrix element
$\langle 1D2\rangle$ is negligible (direct transitions
between levels 1 and 2 are prohibited). In this case
\begin{equation}\label{doubleeins}
{\rho_2\over \rho_1}=
{\hbox{ Re }(g|g)^+_{\omega_2}\over\hbox{ Re }(g|g)^-_{\omega_2}}
{\hbox{ Re }(g|g)^-_{\omega_3}\over\hbox{ Re }(g|g)^+_{\omega_3}}
=
{N(\omega_2)\over N(\omega_2)+1}{N(\omega_3)+1\over N(\omega_3)}
=e^{-\beta(\omega_2)} e^{\beta(\omega_3)}
\end{equation}
Comparing with (\ref{rhooverrho}) and the Einstein emission--absorption
relation we will call this formula the Double Einstein formula.

Relation (\ref{doubleeins}) for the considered system is natural,
since direct transitions from level 2 to level 1 are prohibited
($\langle 1D2\rangle$ is negligible). In this case to jump
from level 2 to level 1 the system have to make two consequent jumps:
from level 2 to level 3 and then from level 3 to level 1.
Therefore it is natural to represent (\ref{doubleeins})
in the following form:
$$
{\rho_1\over \rho_2}=
{W_{\hbox{absorption}}\over W_{\hbox{emission}}}\biggr|_{2-3}
{W_{\hbox{emission}}\over W_{\hbox{absorption}}}\biggr|_{1-3}
=
{N(\omega_2)+1\over N(\omega_2)}{N(\omega_3)\over N(\omega_3)+1}
$$
Note that this formula is true for a special choice of the system
($\langle 1D2\rangle=0$).
Moreover, for a Gibbs state, $\beta$ is linear and
(\ref{doubleeins}) coincides with the Einstein
relation with ${N(\omega_1)+1\over N(\omega_1)}$ at the RHS.

\section{Inverse population state}

Let us consider the following question: for a Gibbs distribution we have
$\rho_1>\rho_2>\rho_3$: the number of particles at an energy level
decreases with
increasing of energy. Can we find such a stationary state where
at least one pair of levels has inversed order: the
number of particles increases with increasing energy?
Such kind of states are important in quantum optics (laser theory).

Let us apply the method of the previous section to
construct a stationary state where $\rho_2>\rho_1$ (population of level 2
is larger than population of level 1).

We will consider the same system considered
in the previous section. In particular, we take
$$\langle 1D2\rangle=0$$
Then from (\ref{doubleeins}) we have $\rho_2>\rho_1$ if and only if
$$
e^{-\beta(\omega_2)} e^{\beta(\omega_3)} >1
$$
This inequality is equivalent to:
\begin{equation}\label{inverse}
\beta(\omega_3)>\beta(\omega_1+\omega_3)
\end{equation}
This means that the local temperature function is non--monotonic
and can decrease with an increase of energy.
Let us note that the quotient $\rho_2/\rho_1$ in the considered approximation
$\langle 1D2\rangle=0$ (metastable level 2) does not depend on
$\langle 1D3\rangle$ and $\langle 2D3\rangle$.
We found that for non--monotonic temperature functions we can have
the inverse population effect: population of level
with higher energy is larger than population of level with lower energy.
In the theory of lasers the inverse population effect sometimes
is discussed as an effect of negative temperature \cite{Haken}.
Indeed if we suppose that the state of the field is equilibrium
and therefore the local temperature function is linear
$\beta(\omega)=\beta\omega$ then (\ref{inverse}) takes the form
$$
\beta\omega_1<0
$$
for $\omega_1>0$.

In our approach we found the inverse population effect  without
introduction of any negative temperature. This effect follows from
the fact that the reservoir is highly non--equilibrium and the
temperature function can decrease with energy.

\section{Master equation for the number operator}

Consider the number operator $n(k)=a^*(k)a(k)$. This operator
has constant free evolution
$$
e^{itH_0} n(k) e^{-itH_0} =n(k)
$$
and therefore does not change in the stochastic limit.
The relation with the master field is as follows
$$
[b_{\omega}(t,k),n(k')]= \lim_{\lambda\to 0}
{1\over \lambda} e^{-{it\over\lambda^2}(\omega(k)-\omega)}[a(k),n(k')]=
$$
$$
=\lim_{\lambda\to 0}
{1\over \lambda} e^{-{it\over\lambda^2}(\omega(k)-\omega)}a(k)\delta(k-k')=
b_{\omega}(t,k)\delta(k-k')
$$
This means that the number operator extends the quantum noise algebra.

Let us find the master equation for the number operator.
Since number operator does not commute with the noises we can not
directly apply the master equation from \cite{notes}.
Applying the QSDE for the evolution operator we get
$$
d\langle U^*_t n(k) U_t\rangle=
\sum_{\omega}\langle U^*_t \biggl(
E_{\omega}^*\left(D\right)dB_{\omega}(t)n(k)
E_{\omega}\left(D\right)dB^*_{\omega}(t)+
$$
$$
+E_{\omega}\left(D\right)dB^*_{\omega}(t)n(k)
E_{\omega}^*\left(D\right)dB_{\omega}(t)
-dt n(k) 2\hbox{Re }G
\biggr)U_t\rangle=
$$
$$
=\sum_{\omega}\langle U^*_t \biggl(
E_{\omega}^*\left(D\right)
E_{\omega}\left(D\right)[dB_{\omega}(t),n(k)]dB^*_{\omega}(t)+
$$
$$
+E_{\omega}\left(D\right)
E_{\omega}^*\left(D\right)[dB^*_{\omega}(t),n(k)]dB_{\omega}(t)
\biggr)U_t\rangle=
$$
$$
=dt\sum_{\omega}\langle U^*_t \biggl(
E_{\omega}^*\left(D\right)
E_{\omega}\left(D\right)
|g(k)|^2 2\pi\delta(\omega(k)-\omega)(N(k)+1)-
$$
$$
-E_{\omega}\left(D\right)
E_{\omega}^*\left(D\right)
|g(k)|^2 2\pi\delta(\omega(k)-\omega)N(k)
\biggr)U_t\rangle
$$
Denoting
$$
\langle X\rangle_t =\langle U^*_t X U_t\rangle
$$
we arrive to master equation
\be\label{n(k)}
{d\over dt}\langle n(k)\rangle_t =
2\pi\sum_{\omega}\delta(\omega(k)-\omega)\biggl(
|g(k)|^2 (N(k)+1)
\langle E_{\omega}^*(D)E_{\omega}(D)\rangle_t-
$$
$$
-|g(k)|^2 N(k)
\langle E_{\omega}(D)E_{\omega}^*(D)\rangle_t\biggr)
\ee
This is a completely general master equation for the number operator
$n(k)$. Therefore to find $\langle n(k)\rangle_t$ it is sufficient
to determine
$\langle E_{\omega}^*(D)E_{\omega}(D)\rangle_t$
and $\langle E_{\omega}(D)E_{\omega}^*(D)\rangle_t$.

Since we consider a generic system by (\ref{1.29c})
equation (\ref{n(k)}) takes the form
\be\label{n(k)1}
{d\over dt}\langle n(k)\rangle_t =
2\pi|g(k)|^2
\sum_{\omega}\delta(\omega(k)-\omega)
|\langle 1_{\omega}| D |2_{\omega}\rangle|^2
\left((N(k)+1)\rho_{2_{\omega}} - N(k)\rho_{1_{\omega}}\right)
\ee

\section{Interaction of atom with radiation in the stationary state}

For a 2--level atom  there is only one term in the $\omega$--summation
in (\ref{n(k)1}) and the stationary state is given by (up to normalization)
$$
\rho_1= \hbox{Re}\,{(g|g)}^-_{\omega},\qquad
\rho_2= \hbox{Re}\,{(g|g)}^+_{\omega}
$$
and therefore the quotient satisfies the Einstein relation:
\be\label{26a}
{\rho_1\over\rho_2}={N(\omega)+1\over N(\omega)}=e^{\beta(\omega)}
\ee

If $N(k)=N(\omega(k))$ for stationary state
for two--level system (\ref{n(k)1}) takes the form
\be\label{2level}
{d\over dt}\langle n(k)\rangle_t =
2\pi\delta(\omega(k)-\omega)|g(k)|^2
|\langle 1_{\omega}| D |2_{\omega}\rangle|^2
\biggl(
N(\omega)(N(\omega)+1)-(N(\omega)+1)N(\omega)
\biggr)=0
\ee
that means that in the stationary state the atom is in
equilibrium with the field. Vanishing of (\ref{2level})
is equivalent to the Einstein relation. Let us note that we
did not assume that the state of the field is equilibrium.

For a 3--level system the master equation for the number of photons
(\ref{n(k)}) takes the form
\be\label{3level}
{d\over dt}\langle n(k)\rangle_t =
2\pi|g(k)|^2\biggl(
\delta(\omega(k)-\omega_1)
|\langle 1|D|2\rangle|^2
\left((N(\omega_1)+1)\rho_{2}-N(\omega_1)\rho_{1}\right)+
$$
$$
+\delta(\omega(k)-\omega_2)
|\langle 1|D|3\rangle|^2
\left((N(\omega_2)+1)\rho_{3}-N(\omega_2)\rho_{1}\right)+
$$
$$
+\delta(\omega(k)-\omega_3)
|\langle 2|D|3\rangle|^2
\left((N(\omega_3)+1)\rho_{3}-N(\omega_3)\rho_{2}\right)\biggr)
\ee
This equation describes the balance of radiation with the 3--level atom.
Consider now the stationary state of the atom. When the
stationary state is equilibrium then in the RHS of (\ref{3level})
each term vanishes separately. This
implies that in an equilibrium state the system is in detailed equilibrium
with the radiation: emission equals to absorption for every frequency.

When the stationary state is non--equilibrium then the Einstein relation
for populations of the levels will be not satisfied. This implies that
each term in (\ref{3level}) will not vanish and the atom will
emit and absorb radiation.

In this case (\ref{x}), (\ref{y}), (\ref{z}) imply
\be\label{term1}
|\langle 1|D|2\rangle|^2((N(\omega_1)+1)\rho_{2}-N(\omega_1)\rho_{1}) =
$$
$$
=|\langle 1|D|2\rangle|^2|\langle 1D3\rangle|^2|\langle 2D3\rangle|^2
I(\omega_2)I(\omega_3)
{e^{\beta(\omega_1)-\beta(\omega_2)+\beta(\omega_3)}-1
\over (e^{\beta(\omega_1)}-1)(1-e^{-\beta(\omega_2)})
(e^{\beta(\omega_3)}-1)}
\ee
\be\label{term2}
|\langle 1|D|3\rangle|^2
\left((N(\omega_2)+1)\rho_{3}-N(\omega_2)\rho_{1}\right)=
$$
$$
=|\langle 1|D|3\rangle|^2 |\langle 1D2\rangle|^2|\langle 2D3\rangle|^2
I(\omega_1)I(\omega_3)
{1-e^{-\beta(\omega_2)+\beta(\omega_1)+\beta(\omega_3)}
\over (1-e^{-\beta(\omega_2)})(e^{\beta(\omega_1)}-1)
(e^{\beta(\omega_3)}-1)}
\ee
\be\label{term3}
|\langle 2|D|3\rangle|^2
\left((N(\omega_3)+1)\rho_{3}-N(\omega_3)\rho_{2}\right)=
$$
$$
=|\langle 2|D|3\rangle|^2 |\langle 1D2\rangle|^2|\langle 1D3\rangle|^2
I(\omega_1)I(\omega_2)
{e^{\beta(\omega_3)+\beta(\omega_1)-\beta(\omega_2)}-1
\over (e^{\beta(\omega_3)}-1)(e^{\beta(\omega_1)}-1)
(1-e^{-\beta(\omega_2)})}
\ee
Note that (\ref{term1}), (\ref{term2}), (\ref{term3}) contain
the combination $\beta(\omega_3)+\beta(\omega_1)-\beta(\omega_2)$
that vanishes for an equilibrium state of the reservoir.

Equation (\ref{3level})  takes the form
\be\label{3level1}
{d\over dt}\langle n(k)\rangle_t =
2\pi|g(k)|^2
|\langle 1|D|2\rangle|^2|\langle 1D3\rangle|^2|\langle 2D3\rangle|^2
{e^{\beta(\omega_1)-\beta(\omega_2)+\beta(\omega_3)}-1
\over (e^{\beta(\omega_1)}-1)(1-e^{-\beta(\omega_2)})
(e^{\beta(\omega_3)}-1)}
$$
$$
\biggl(I(\omega_2)I(\omega_3)\delta(\omega(k)-\omega_1)-
I(\omega_1)I(\omega_3)\delta(\omega(k)-\omega_2)
+I(\omega_1)I(\omega_2)\delta(\omega(k)-\omega_3)
\biggr)
\ee
Equation (\ref{3level1}) shows that the 3--level system in non--equilibrium
stationary state converts radiation with
the energy $\omega_2=\omega_1+\omega_3$ into radiation with energies
$\omega_1$ and $\omega_3$ if
$$
\beta(\omega_1)+\beta(\omega_3)>\beta(\omega_1+\omega_3)
$$
and vice versa in the case of the opposite inequality.

Integrating (\ref{3level1}) over $k$ we get
\be\label{ddtofE}
{d\over dt}\int E(\omega(k))\langle n(k)\rangle_t dk=
2\pi |\langle 1|D|2\rangle|^2|\langle 1D3\rangle|^2|\langle 2D3\rangle|^2
$$
$$
I(\omega_1)I(\omega_2)I(\omega_3)
\left(E(\omega_1)-E(\omega_2)+E(\omega_3)\right)
{e^{\beta(\omega_1)-\beta(\omega_2)+\beta(\omega_3)}-1
\over (e^{\beta(\omega_1)}-1)(1-e^{-\beta(\omega_2)})(e^{\beta(\omega_3)}-1)}
\ee

When $E(\omega)$ is the dispersion of the field
$$
E(\omega)=\omega
$$
then
$$
E(\omega_1)-E(\omega_2)+E(\omega_3)=0
$$
and (\ref{ddtofE}) implies the conservation of energy
$$
{d\over dt}\int \omega(k)\langle n(k)\rangle_t dk =0
$$
For the time derivative of number operator we get
$$
\int {d\over dt}\langle n(k)\rangle_t dk=
2\pi |\langle 1|D|2\rangle|^2|\langle 1D3\rangle|^2|\langle 2D3\rangle|^2
$$
$$
I(\omega_1)I(\omega_2)I(\omega_3)
{e^{\beta(\omega_1)-\beta(\omega_2)+\beta(\omega_3)}-1
\over (e^{\beta(\omega_1)}-1)(1-e^{-\beta(\omega_2)})(e^{\beta(\omega_3)}-1)}
$$
Using that $\omega_2=\omega_1+\omega_3$ we obtain that if
\be\label{absorps}
\beta(\omega_1)+\beta(\omega_3)<\beta(\omega_1+\omega_3)
\ee
then the derivative is negative and the system absorbs the radiation
(total number of absorbed photons is larger than the total
number of emitted photons).
In this case (\ref{3level1}) implies that
the system absorbs photons with frequencies $\omega_1$ and $\omega_3$
and emits photons with frequency $\omega_2$.

If
\be\label{emits}
\beta(\omega_1)+\beta(\omega_3)>\beta(\omega_1+\omega_3)
\ee
then the derivative is positive and the system emits the radiation
(the total number of absorbed photons is smaller than the total
number of emitted photons).
In this case (\ref{3level1}) implies that
the system emits photons with frequencies $\omega_1$ and $\omega_3$
and absorbs photons with frequency $\omega_2$.
For instance in the case of inverse population (\ref{inverse})
the condition of emission regime (\ref{emits}) is satisfied.

These regimes of emission and absorption
are controlled only by the difference
$\beta(\omega_1)+\beta(\omega_3)-\beta(\omega_1+\omega_3)$.

\bigskip

\centerline{\bf Acknowledgements}
The authors are grateful to I.V.Volovich for discussions.
Kentaro Imafuku and Sergei Kozyrev are grateful to Centro Vito Volterra
and Luigi Accardi for kind hospitality. This
work was partially supported by INTAS 9900545 grant.
Kentaro Imafuku  is supported by an overseas research fellowship
of Japan Science and Technology Corporation.
Sergei Kozyrev was partially supported by RFFI 990100866 grant.

\end{document}